\documentclass[structabstract]{aa} 
\usepackage{graphicx}
\usepackage{natbib}
\usepackage{url}
\usepackage[czech,english]{babel}

\usepackage[utf8x]{inputenc}

\usepackage{amsmath}
\usepackage{layouts}
\usepackage{color}
    \definecolor{Blue}{rgb}{0.0,0.0,1.0}
    \definecolor{Red}{rgb}{1.0,0.0,0.0}
    \definecolor{Green}{rgb}{0.0,1.0,0.0}

\newcommand{\be}{\begin{equation}}                                 
\newcommand{\ee}{\end{equation}}                                   
\newcommand{\bea}{\begin{eqnarray}}                                
\newcommand{\eea}{\end{eqnarray}}  
\newcommand{\bi}{\begin{itemize}}                                 
\newcommand{\ei}{\end{itemize}}                                      
                                                   
\definecolor{gray}{rgb}{.6,.6,.6}                                  %
\definecolor{green}{rgb}{0,.6,0}                                   %
\definecolor{red}{rgb}{0.6,0,0}                                    %

\begin{document}
\title{Twin peak HF QPOs as a spectral imprint of dual oscillation modes of accretion tori}
%
\author{      P. Bakala\inst{1}
\and           K. Goluchov{\'a}\inst{1}
\and           G. T{\"o}r{\"o}k\inst{1}
\and           E. {\v S}r{\'a}mkov{\'a}\inst{1}
\and           M. A. Abramowicz\inst{1, 2, 4}
\and           F. H. Vincent\inst{2}
\and           G. P. Mazur\inst{2, 3}
}
\institute{
Institute of Physics, Faculty of Philosophy and Science, Silesian University in Opava, Bezrucovo nam. 13, CZ-746 01 Opava, Czech Republic
									\\ \email{pavel.bakala@fpf.slu.cz}
									\\ \email{katka.g@seznam.cz}
                  \\ \email{gabriel.torok@gmail.com}
\and             
              Nicolaus Copernicus Astronomical Center, ul. Bartycka 18, PL-00-716 Warszawa, Poland
	   \\ \email{fvincent@camk.edu.pl}
\and    
Institute of Physics, Polish Academy of Sciences, aleja Lotnikow 32/46, PL-02-668 Warszawa, Poland     
                 \\ \email{gmazur@camk.edu.pl}
\and          Physics Department, Gothenburg University,
               SE-412-96 G{\"o}teborg, Sweden
                 \\ \email{marek.abramowicz@physics.gu.se}
}
   \date{Received ; accepted }
\abstract 
   {High frequency (millisecond) quasi-periodic oscillations (HF QPOs) are observed in the X-ray power-density spectra of several microquasars and low mass X-ray
binaries. Two distinct QPO peaks, so-called twin peak QPOs, are often detected simultaneously exhibiting their frequency ratio close or equal to 3/2.
A widely discussed class of proposed QPOs models is based on oscillations of accretion toroidal structures orbiting in the close vicinity of black holes or neutron stars. 
	}
	 {Following the analytic theory and previous studies of observable spectral signatures, we aim to model the twin peak QPOs as a spectral imprint of specific dual oscillation regime defined by a combination of the lowest radial and vertical oscillation mode of slender tori. We consider the model of an optically thick slender accretion torus with constant specific angular momentum. We examined power spectra and fluorescent $K\alpha$ iron line profiles for two different simulation setups with the mode frequency relations corresponding to the epicyclic resonance HF QPOs model and modified relativistic precession QPOs model.
}
   {We use relativistic ray-tracing implemented in parallel simulation code LSDplus. In the background of the Kerr spacetime geometry, we analyze the influence of the distant observer inclination and the spin of the central compact object. Relativistic optical projection of the oscillating slender torus is illustrated by images in false colours related to the frequency shift.
}
   {We show that performed simulations yield power spectra with the pair of dominant peaks corresponding to the frequencies of radial and vertical oscillation modes with the proper ratio equal to $3/2$ on a wide range of inclinations and spin values. We also discuss exceptional cases of a very small and very high inclination as well as unstable high spin relativistic precession-like configuration predicting constant frequency ratio equal to $1/2$. We demonstrate significant dependency of broadened $K\alpha$ iron line profiles on the inclination of the distant observer.}  
   {This study presents a further step towards the proper model of oscillating accretion tori producing HF QPOs. More realistic future simulations should be based on incorporation of the resonant coupling of oscillation modes, influence of torus opacity and pressure effects on the mode frequencies and the torus shape.
   }

\authorrunning{Bakala et al.}
\titlerunning{Twin peak HF QPO as a imprint of dual oscillation modes}
\maketitle



\section{Introduction}

High frequency quasi-periodic oscillations (HF QPOs) have been observed in several microquasars and low mass X-ray binaries (LMXBs). Their frequencies are roughly comparable to frequencies of orbital motion of test particles in the vicinity of the central compact object. 
The black hole HF QPOs occur at frequencies that are characteristic for a particular source and seem to be constant in time. Strictly speaking, the current observational data enable to distinguish HF QPOs only for fairly specific spectral states \citep{bel-san-men:2012}. However, provided that HF QPO with lower amplitudes and different frequencies exist in the remaining time, their detection goes beyond our present technological capabilities.

Two kinds of sharp HF QPO peaks can be distinguished, so-called lower and upper HF QPOs \citep[see] [for a review]{kli:2004,Rem-MC:2006}. If both HF QPOs peaks are observed simultaneously (twin-peak HF QPOs), the ratio of frequencies of upper and lower HF QPO peaks is often close to 3:2 indicating the possible presence of unspecified resonant phenomena  \citep{abr-klu:2001, tor-etal:2005} \footnote{In the case of LMXBs, there are indications that the ratio of twin peak HF QPOs frequencies is clustered not only around 3:2, but also 4:3 and 5:4 ratios are observed \citep[see, e.g.,][]{tor-etal:2007, tor-etal:2008b, tor-etal:2008c,tor-etal:2008a}.}.
At present, there is no consensus on the QPO nature. However, the inverse mass scaling of the QPOs frequencies \citep{abr-klu:2001}  provides a strong argument for interpreting the observed QPO peaks by the frequencies of perturbed orbital motion in the strong gravitational field or oscillation of some accretion structures.
accretion structures. 
Such interpreting naturally promises an attractive possibility to measure the mass and spin of the black hole \citep[e.g.][]{abr-klu:2001,abr-fra:2013,ing-mot:2014,mot-etal:2014,tor:2005,Tor-etal:2012}.
Many QPOs models were proposed \citep[e.g.][and others]{alp-sha:1985, lam-etal:1985, mil-etal:1998a,psa-etal:1999b,Ste-Vi:1999,abr-klu:2001,abr-klu:2004,Kat:2001,tit-ken:2002,rez-etal:2003,sch-ber:2004,pet:2005a,zha:2005,tak-var:2006,kat:2007,stu-etal:2008,muk:2009,cad-etal:2008,kos-etal:2009,ger-etal:2009,ger:2013,lai-etal:2013,pech-etal:2008,pech-etal:2013}, however each model still faces several difficulties. Moreover, the capabilities of the present X-ray observatories (e.g., Rossi X-ray Timing Explorer -- RTXE) are insufficient to adequately analyze the harmonic content of the power spectra of observed lightcurves, which can be crucial to distinguishing between particular QPO models \citep{bak-etal:2014a,kar-etal:2014}. 
Hopefully, the proposed future instruments represented by e.g. Large Observatory for X-ray Timing (LOFT) project \citep{loft:2014}, targeted to explore strong gravity environment, will advance QPOs observational possibilities.

A specific class of QPO models assumes oscillations excited in accretion tori. 
The first QPO model involving numerically modelled thick accretion tori was developed by \cite{rez-etal:2003} and related light curves and power spectra were analyzed by \cite{sch-rez:2006}. Light curves and power spectra of radially and vertically oscillating slender torus with a circular cross-section were investigated in the background of the Schwarzschild geometry by \cite{bur-etal:2004}. Numerical simulations of epicyclic modes of tori oscillations were compared to the analytical results by \cite{sra:2005}. Later on, there has arose of studies devoted to more realistic analytic treatment of oscillating slender tori \citep[e.g.][]{abr-etal:2006,bla-etal:2006}. Such studies were extended to the case of non-slender tori by \cite{str-sra:2009}. 

In this paper, we examine timing properties of the numerically simulated flux emitted from a slender, polytropic, perfect fluid, non-self-gravitating accretion torus with constant specific angular momentum, which oscillates simultaneously in radial and vertical directions. \footnote{The slender torus geometry considered here is in good accord with the truncated disk model \citep[see, e.g.,][]{don-gie-kub:2007}. Also, the power spectral fits of \cite{ing-don:2012} predict that the hot inner flow (corona) has a small scale height in the relevant spectral state.} The article is a follow-up of the studies of \cite{maz-etal:2013}  and \cite{vin-etal:2014} who investigated the observable signatures of simple time-periodic slender torus deformations as well as slender torus motion described by the set of  oscillation modes derived by \cite{bla-etal:2006}. Using their results, we took the next step towards a fully realistic model of HF QPO based on oscillations of accretion tori. In order to model simultaneously detected twin peaks HF QPOs pairs, we define a new dual oscillation mode as a linear combination of two lowest slender torus oscillation modes: radial and vertical ones. We assume that the observed twin peaks HF QPOs can be identified with the pairs of the most prominent peaks in the modelled power spectra.  We examine two different setups of the dual oscillation regime corresponding to the twin peaks HF QPOs frequency relation of epicyclic resonance HF QPOs model  \citep{abr-klu:2001} and slightly modified analogous relations of relativistic precession QPOs model \citep{Ste-Vi:1999,Tor-etal:2012}. Those two competing QPOs models are probably the most discussed ones at present \citep{loft:2014}.

Our simulations are performed in the background of the Kerr geometry corresponding to the case of microquasars with the black hole binary component. It was shown that the spacetime around slowly rotating high-mass neutron stars can be fairly well approximated by the Kerr metric \citep{Tor-etal:2010,Tor-etal:2012,urb-etal:2013}. Therefore, such a finding sets conditions and constrains for the applicability of presented results for the case of LMXBs with neutron star. 

The commonly accepted model of X-ray energy spectrum of microquasars and LMXBs assumes that the illumination of the cold accretion disk or torus by the primary component of X-ray spectrum, interpreted as the inverse Compton scattering of thermal photons in a hot corona, produces spectral lines by fluorescence. 
The strongest observed line is the $K\alpha$ iron line located at $6.4\,keV$ in the rest frame. Observed broad profiles of the spectral lines are substantially influenced by the spacetime metric, the geometry of the emitting region and the distant observer inclination \citep{fab-etal:1989,cad-cal:2005,bam:2013}.  In order to develop an additional tool for distinguishing various configurations of radiating slender torus, we compute related $K\alpha$ iron line profiles.

The article consits of the following parts:
Section~\ref{sec:torus} is devoted to the theory which describes the slender torus model and its dual oscillation regime.
Section~\ref{sec:torusmodel} provides more details of the investigated particular setup of the model of slender torus.
Section \ref{sec:raytracing} describes our numerical implementation of relativistic ray-tracing and the following construction of lightcurves, power spectra  and iron $K\alpha$ line profiles.
Section \ref{sec:methods} is devoted to the methodology of simulations performance and results analysis.
Section~\ref{sec:results} shows the obtained results, and Section~\ref{sec:conc} gives conclusions and future research perspectives

\section{Slender torus}
\label{sec:torus} 

\subsection{Equilibrium torus configuration}

We consider an axisymmetric, non self-gravitating, perfect fluid, constant specific angular momentum, circularly orbiting accretion torus in the background of the Kerr geometry. Using the  $(-+++)$ signature and geometrical units  ($c = G = M = 1$), the line element of the Kerr spacetime in Boyer-Lindquist coordinates parameterized by specific angular momentum (spin) $a$ reads
\begin{eqnarray}
 \mathrm{d}s^2 &=& -\left(1-\frac{2  r}{\Sigma}\right) \,\mathrm{d}t^2 
  - \frac{4 r a }{\Sigma} \sin^2\theta\,\mathrm{d}t\, \mathrm{d}\varphi 
  + \frac{\Sigma}{\Delta}\, \mathrm{d}r^2  \nonumber \\
&+& \Sigma \,\mathrm{d}\theta^2
  + \left(r^2 + a^2 +\frac{2 r a^2 \sin^2\theta}{\Sigma}\right)\sin^2\theta\, \mathrm{d}\varphi^2, 
\label{kerr_metric}
\end{eqnarray}
where $\Sigma \equiv r^{2} + a^{2}\cos^{2}\theta$ and $\Delta \equiv r^{2} - 2r + a^{2}$. 
Moreover, we assume that the radial extent of the torus cross-section is small compared to its central radius. The perfect fluid forming the torus is described by polytropic equation of state (with a polytropic constant $K$ and polytropic index $n$). In such a case, the energy density $e$ is a function of the pressure $p$ and the mass density $\rho$. The pressure is a function of mass density only. The torus surface is given by the zero pressure, while the pressure gradient is equal to zero in the torus centre. 
Using the conservation law, one can get \citep[see][]{abr-etal:2006,bla-etal:2006}
\begin{equation}
\frac{p}{\rho} = \frac{p_{0}}{\rho_{0}}f(r,\theta),
\end{equation}
where the surface function $f$ is constant at isobaric and
isodensity surfaces. Naturally, the surface function vanishes at the torus surface.
Here and below, values of the quantities denoted with a subscript $_0$ are taken in the centre of the torus in the equilibrium state.

Following \cite{abr-etal:2006}, we use the new radial and vertical coordinates 
\begin{equation}
\bar{x} = \left( \sqrt{g_{rr}} \right)_0 \left( \frac{r -
r_0}{ \beta r_0}\right),~~
\bar{y} = \left( \sqrt{g_{\theta\theta}} \right)_0 \left( \frac{\pi/2 -
\theta}{\beta r_0}\right).
\label{cross-coordinates}
\end{equation}
with zero in the torus centre. Parameter $\beta$ determines the torus thickness and is given by formula 
\begin{equation}
\beta^2 = \frac{2n c^2_{s0}}{r^2_0 (u^{t}_{0})^2 \Omega^2_0}\,,
\label{eq:beta}
\end{equation}
where $u^{t}_{0}$ denotes the four-velocity of the fluid and square of the sound speed $c_{s0}^{2} << 1$ can be expressed in the form
\begin{equation}
c_{so}^2 = \left(\frac{\partial p}{\partial \rho}\right)_{0} = \frac{n + 1}{n} \frac{p_{0}}{\rho_{0}}.
\end{equation}
Keplerian angular velocity $\Omega_{K0}$ in the Kerr spacetime reads
\begin{equation}
\Omega_{K0} = 1/(r_{0}^{3/2} + a)\,. \label{OMEGA}
\end{equation}
In the case of slender torus, the parameter $\beta$ must fulfil the condition
\begin{equation}
\beta \ll 1\,. \label{slendercondition}
\end{equation}
In such a coordinate frame, the surface function $f(r, \theta)$ can be rewritten as
\begin{equation}
f(\bar{x},\bar{y}) = 1 - \bar{\omega}_{r}^{2}\bar{x}^{2} - \bar{\omega}_{\theta}^{2}\bar{y}^{2}.
\end{equation}
where 
\begin{eqnarray}
\bar{\omega}_{r} &=& \sqrt{1 - \frac{6}{r_{0}} + \frac{8a}{r_{0}^{3/2}} - \frac{3a^{2}}{r_{0}^{2}}}\,, \label{omegar} \\
\bar{\omega}_{\theta} &=& \sqrt{1 - \frac{4a}{r^{3/2}_{0}} +  \frac{3a^{2}}{r_{0}^{2}}}\,, \label{omegatheta} 
\end{eqnarray}
are radial and vertical epicyclic frequencies of free test particles in the centre of the torus scaled to $\Omega_{K0}$ \citep[see, e.g.,][]{ali-gal:1981,abr-klu:2005,ali:2008}.
We can easily see, that the slender torus cross-section (given by $f(\bar{x},\bar{y}) = 0$) has elliptical shape in the $\bar{x}$-$\bar{y}$ plane.

\begin{figure*}
\centering
	 \includegraphics[width=1.0\hsize]{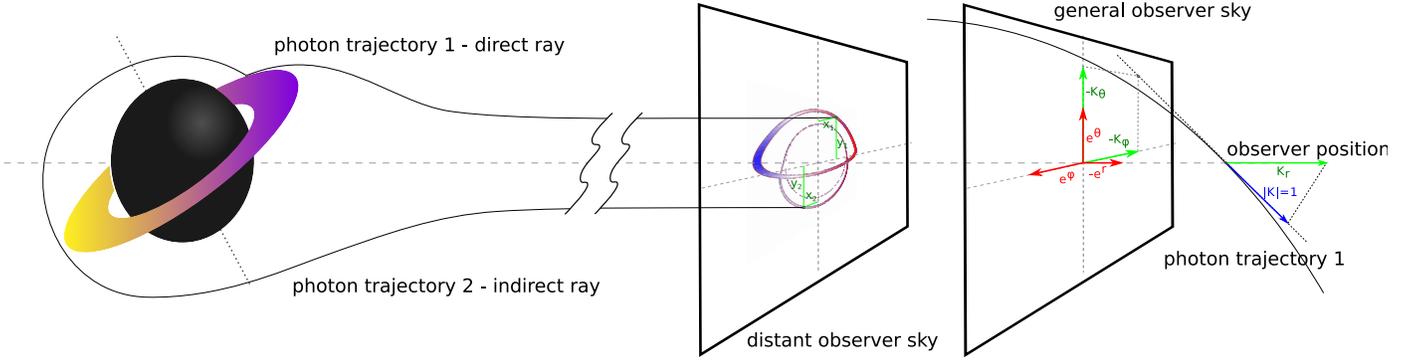}
  \caption{Schematic ray-tracing geometry for the cases of a general observer and distant observer ($r_{obs}\to\infty$).}
  \label{fig_5}
\end{figure*}

\subsection{Oscillating slender torus}

Let us assume small pressure perturbation in the form
\begin{equation}
\delta p \propto  e^{i(m\varphi - \Omega t)}.
\end{equation}
where $m$ is azimuthal wave number ($m\in \mathbf{N}$) and $\Omega$ is oscillation angular frequency. 
It is useful to introduce a new variable  - eigenfuction $W_i$, which is the function of perturbation in pressure in the form
\begin{equation}
W_i = -\frac{\delta p}{u^{t}_{0}\rho\sigma_i}, 
\end{equation}
where related eigenfrequency $\sigma_i$ reads
\begin{equation}
\sigma_i=\Omega_i-m_i\Omega_{K0}.
\end{equation}
The perturbative surface equation $\tilde{f}(r,\theta)$ and torus four-velocity  can be expressed in the form \citep{vin-etal:2014}
\begin{eqnarray}
  \tilde{f}(r,\theta) &=& f(r,\theta) - \frac{1}{n + 1}\frac{\rho_{0}}{p_{0}}u^{t}_{0}Re\left\{W_i\right\} \sigma_i, \label{surfaceeq}\\
	u^{\mu} &=& u^{\mu}_{0} + Re\left\{\frac{i\rho_{0}}{p_{0} + e_{0}}\left(\frac{\partial W_i}{\partial x^{\mu}}\right)\right\}.
\end{eqnarray}
Discrete set of eigefunctions $W_i$ and related eigenfrequencies $\sigma_i$ describing different oscillation modes was found by \cite{bla-etal:2006} solving  Papaloizou-Pringle equation for slender torus case corresponding to the condition $\beta \rightarrow 0$ \citep{abr-etal:2006,pap-pri:1984}. Here we use only two simplest solutions: radial oscillation mode and vertical oscillation mode. 
In the case of radial oscillation mode, the eigenfuction and related eigenfrequency are given as
\begin{equation}
W_{r} = a_{r} \bar{x}e^{i(m_r\varphi - \Omega_r t)}\,,\qquad \sigma_{r} = \bar{\omega}_{r}\Omega_{K0}\,,
\end{equation} 
where $a_{r}$ is a free parameter related to the amplitude of oscillations.
Then the surface equation (\ref{surfaceeq}) takes the form
\begin{equation}
1 - \bar{\omega}_{r}^{2}\bar{x}^{2} - \bar{\omega}_{\theta}^{2}\bar{y}^{2} 
  - A_{r}\bar{x} \cos(m_{r}\varphi - \Omega_{r}t) = 0, \label{rsurfaceeq}
\end{equation} 
where amplitude $A_{r}$ is given by the formula 
\begin{equation}
A_{r} = \frac{1}{n+1}\frac{\rho_{0}}{p_{0}}u^{t}_{0}\sigma_{r} a_{r}\,,
\end{equation}
and mode oscillation angular frequency reads
\begin{equation}
\Omega_{r}=\sigma_{r} + \Omega_{K0} m_{r}\,. \label{rad.os.f}
\end{equation}
In order to rewrite the surface equation (\ref{rsurfaceeq}) to the form
\begin{equation}
1 - \bar{\omega}_{r}^{2}\left(\bar{x} + \delta \bar{x}\right)^{2} - \bar{\omega}_{\theta}^{2}\bar{y}^{2}= 0\,, \label{rad.os}
\end{equation}
the term
\begin{equation}
\left(\frac{A_{r}}{2\bar{\omega}_{r}} \cos(m_{r}\varphi - \Omega_{r}t)\right)^{2}
\end{equation}
must be added.
Then the displacement $\delta \bar{x}$ can be expressed by the formula
\begin{equation}
\delta \bar{x} = \frac{A_{r}}{2\bar{\omega}_{r}^{2}} \cos\left(m_{r}\varphi - \Omega_{r}t\right). \label{rad.disp}
\end{equation}
In this approximation, the added term should be small enough, which corresponds to the condition for radial amplitudes
\begin{equation}
\frac{A_{r}^{2}}{4\bar{\omega}_{r}^{2}} << 1. \label{r_small_ampl}
\end{equation}
The radial component of the surface four-velocity is simply given by a derivative of the displacement $\delta \bar{x}$ with respect to proper time $\tau$ as  
\begin{equation}
u^{r}_{sur} =\frac{dr}{d\tau} = \frac{d\bar{x}}{d\tau} \frac{\beta r_{0}}{\left( \sqrt{g_{rr}} \right)_0} = \frac{d(-\delta \bar{x})}{d\tau} \frac{\beta r_{0}}{\left( \sqrt{g_{rr}} \right)_0}\,,
\end{equation}
and it takes the covariant form
\begin{equation}
u_{r}^{sur} = - \beta r_{0}\left( \sqrt{g_{rr}} \right)_0\frac{A_{r}}{2\bar{\omega}_{r}}\Omega_{K0}u^{t}_{0}\sin(m_{r}\varphi - \Omega_{r}t). \label{rad.os.u}
\end{equation}
Analogously, in the case of the vertical oscillation mode, the eigenfunction and related eigenfrequency are given as
\begin{equation}
W_{\theta} = a_{\theta} \bar{y}e^{i(m_{\theta}\varphi - \Omega_{\theta} t)}\,,\qquad \sigma_{\theta} = \bar{\omega}_{\theta}\Omega_{K0}\,.
\end{equation} 
Using the condition for vertical  amplitudes $\frac{A_{\theta}^{2}}{4\bar{\omega}_{\theta}^{2}} << 1$, the surface equation, displacement $\delta \bar{y}$ and surface four-velocity component $u_{\theta}^{sur}$ for the vertical mode can be expressed as 	
\begin{eqnarray}	
  &&1 - \bar{\omega}_{r}^{2}\bar{x}^{2} - \bar{\omega}_{\theta}^{2}\left(\bar{y} + \delta \bar{y}\right)^{2}= 0\,, \label{ver.os} \\
	&&\delta \bar{y} = \frac{A_{\theta}}{2\bar{\omega}_{\theta}^{2}} \cos\left(m_{\theta}\varphi - \Omega_{\theta}t\right)\,, \label{ver.disp}\\
 &&u_{\theta}^{sur} =  \beta r_{0}\left(\sqrt{g_{\theta \theta }}\right)_0 \frac{A_{\theta}}{2\bar{\omega}_{\theta}}\Omega_{K0}u^{t}_{0}\sin(m_{\theta}\varphi - \Omega_{\theta}t)\,, \label{ver.os.u}
\end{eqnarray}
where amplitude $A_{\theta}$ given by formula 
\begin{equation}
A_{\theta} = \frac{1}{n+1}\frac{\rho_{0}}{p_{0}}u^{t}_{0}\sigma_{\theta}a_{\theta}\,,
\end{equation}
and mode oscillation angular frequency reads
\begin{equation}
\Omega_{\theta}=\sigma_{\theta} + \Omega_{K0} m_{\theta}\,. \label{ver.os.f}\\
\end{equation}

\subsection{Dual oscillation mode}

In such an approximation, radial and vertical oscillations are independent, and we can easily combine them into a new surface equation
\begin{equation}
  1 - \bar{\omega}_{r}^{2}\left(\bar{x} + \delta \bar{x}\right)^{2} - \bar{\omega}_{\theta}^{2}\left(\bar{y} + \delta \bar{y}\right)^{2}= 0. \label{dualsurface}
\end{equation}
The equation above defines the dual oscillation mode with four parameters: amplitudes $A_{r}$, $A_{\theta}$ and azimuthal wave numbers $m_{r}$, $m_{\theta}$. 

\section{Investigated model of oscillating slender torus}
\label{sec:torusmodel} 

\subsection{Location, thickness and frequencies identification}

Radial and vertical angular frequencies $\Omega_{r},\Omega_{\theta}$ of the dual oscillation mode are given as a linear combination of eigenfrequencies $\sigma_{r},\sigma_{\theta}$ and Keplerian angular velocity $\Omega_{K0}$. Therefore the location of the torus centre in equilibrium $r_{0}$, value of the spin of the central Kerr black hole and wave number pair $m_{r}$, $m_{\theta}$ fully determine the ratio of oscillation  angular frequencies $\Omega_{\theta}/\Omega_{r}$. Considering the properties of epicyclic and orbital frequencies in the Kerr spacetime \citep{tor-stu:2005}, our model identifies the lower and upper kHz QPOs frequencies with $\nu_l=\left|\Omega_{r}/2\pi\right|$ and $\nu_u=\left|\Omega_{\theta}/2\pi\right|$, respectively. As the aim of the article is to model twin peak HF QPOs  with peaks frequency ratio close to $3:2$, we choose $r_{0}$ in such a way, that the ratio $\Omega_{\theta}/\Omega_{r}$ is just equal to $3:2$ for a given wave number pair. \footnote{Investigated torus model does not consider the presence of the resonant coupling of oscillation modes, which can cause amplification or excitation of QPOs on preferred values of radial coordinate \citep{hor:2008}.} In each of such positions of the torus centre we set the radial extent of the torus cross-section to $r_{0}$/10. Corresponding values of the parameter $\beta$ are in accordance with slender torus condition (\ref{slendercondition}). Arbitrary amplitudes of oscillations are fixed by setting $A_{r,\theta}=\bar{\omega}_{r,\theta}$.

\subsection{Optical properties of the torus}
The torus described above is assumed to be optically thick.
Moreover, we use two other very simple assumptions. The torus surface emits radiation isotropically in its comoving local frame and the local flux integrated over the surface area of a thin vertical slice of the torus is conserved. Such a surface area is proportional to $r_c(t_{em},\phi)\, \times \, C(t_{em},\phi)$, where $C(t_em,\phi)$ is the torus cross-section circumference, $t_{em}$ is the time of emission and $r_c(t_{em},\phi)$ is the radial coordinate of the centre of the torus cross-section. In the case of the used approximation, the investigated dual oscillation mode describes pure radial and vertical displacements, and the torus cross-section circumference remains constant. Therefore, using normalisation to $1$ for the equilibrium state, the local emitted intensity simply reads  
\begin{equation}
I_{em}(t_{em},\phi)=r_0/r_c(t_{em},\phi). \label{local_intensity}
\end{equation}

\subsection{Epicyclic resonance axisymmetric setup}
The first investigated torus setup combines pure epicyclic axisymmetric vertical and radial
oscillation, where radial and vertical oscillation frequencies are identical with radial $\nu_r=\bar{\omega}_{r}\Omega_{K0}/2\pi$ and vertical $\nu_{\theta}=\bar{\omega}_{\theta}\Omega_{K0}/2\pi$ epicyclic frequencies, respectively. This setup corresponds to the often quoted epicyclic resonance  (ER) HF QPOs model  \citep{abr-klu:2001} based on the presence of non-linear resonant phenomena between epicyclic disc oscillation modes \citep{klu-abr:2001,abr-etal:2003b,abr-etal:2003c,hor:2008}.  
The dual oscillation mode with such behaviour is related to wave number pair $m_{r}=0$, $m_{\theta}=0$. Therefore the frequency relation determining the radial coordinate of the torus centre $r_{0}$ reads
\begin{equation}
\frac{\nu_u}{\nu_l}\Bigg\arrowvert_{m_{r}=0}^{m_{\theta}=0}=\frac{\nu_{\theta}}{\nu_{r}}=\frac{3}{2}. \label{model1rel}
\end{equation}

\subsection{Relativistic precession-like non-axisymmetric setup}

The dual oscillation mode related to wave number pair $m_{r}=-1$, $m_{\theta}=-2$ yields the frequency relation in the form 
\begin{equation}
\frac{\nu_u}{\nu_l}\Bigg\arrowvert_{m_{r}=-1}^{m_{\theta}=-2}=\frac{2\nu_K-\nu_{\theta}}{\nu_K-\nu_{r}}=\frac{3}{2}\,, \label{model2rel}
\end{equation}
where $\nu_K=\Omega_{K0}/2\pi$ is the Keplerian orbital frequency at $r_{0}$. As the denominator matches the periastron precession frequency, in the Schwarzschild case the relation corresponds exactly to the relation of the relativistic precession (RP) QPOs model
\begin{equation}
\frac{\nu_u}{\nu_l}\Bigg\arrowvert_{RP}=\frac{\nu_K}{\nu_K-\nu_{r}}\,. \label{RPrel}
\end{equation}
The RP model was proposed in a series of papers by \cite{ste-vie:1998,Ste-Vi:1999,sta-vie:2002,mor-ste:1999} and explains the QPOs as a direct manifestation of relativistic epicyclic motion of radiating blobs \citep{Ste-Vi:1999}.
In the case of slow rotation the frequency relation (\ref{model2rel}) still almost coincides with the ratio of twin peaks HP QPOs frequencies predicted by the relation (\ref{RPrel})  \citep{Tor-etal:2012}.
Such an approach can be understood as a redefinition of the modulation mechanism of RP model, however, it preserves predictive power of the model.

\section{Numerical modelling of radiation emission, propagation and detection}
\label{sec:raytracing}

\begin{figure*}
\centering
\includegraphics[width=1.0\hsize]{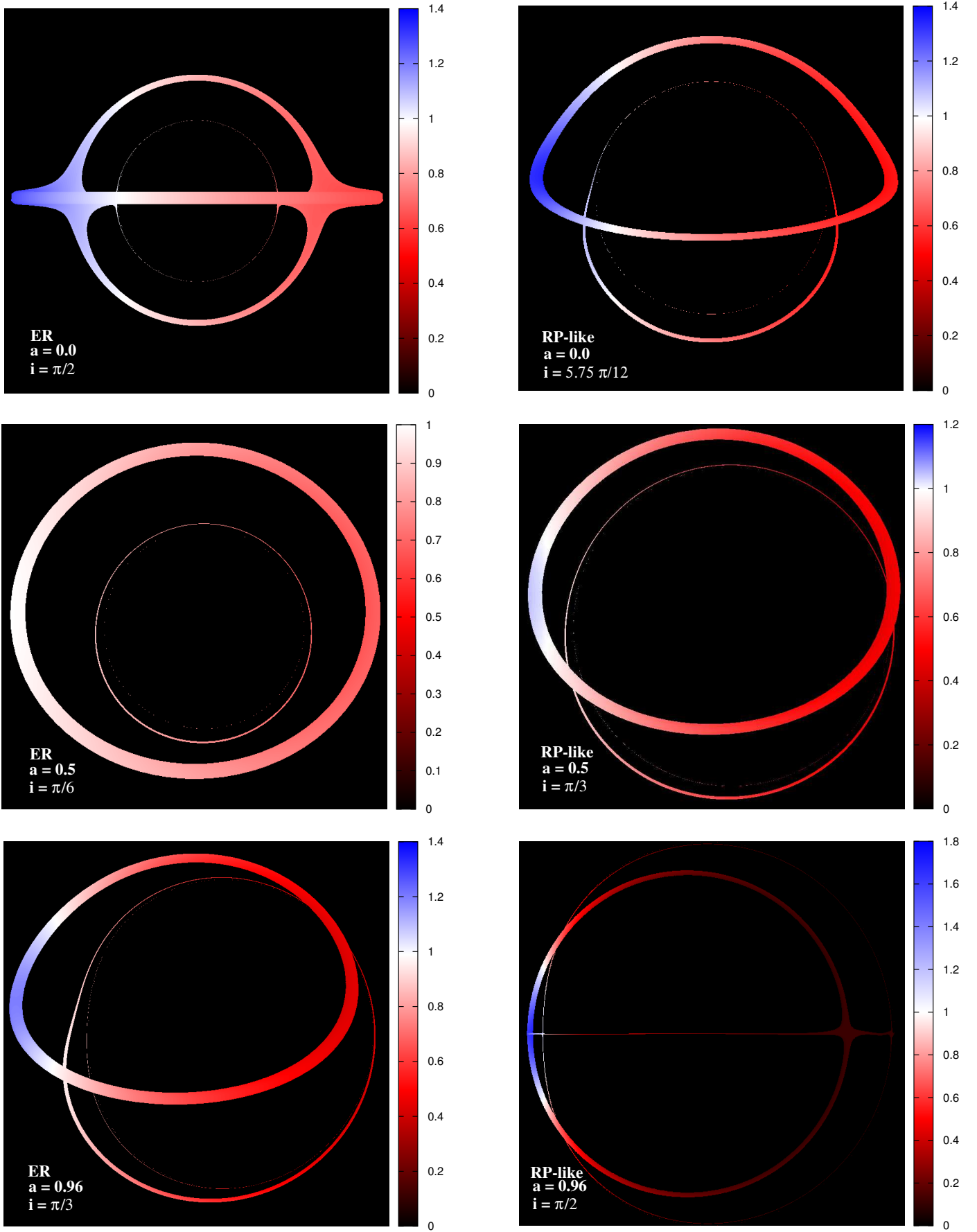}
\caption{\label{FSmaps}Examples of the frequency shift maps for different distant views of the oscillating slender torus. The colour boxes on the right display the false colours scale of frequency shift values. The left column corresponds to the case of the ER axisymmetric dual mode. The right column corresponds to the case of the RP-like non-axisymmetric dual mode. The rows of the figure correspond to spin values $a=0$,\,$a=0.5$ and $a=0.96$, respectively. }
\end{figure*}

\subsection{Ray-tracing in the Kerr spacetime}

Relativistic ray-tracing is a key ingredient of proper models of relativistic imaging, lightcurves of accretion structures, related power spectra as well as relativistic spectral line profiles. Different ray-tracing techniques were developed using direct numerical integration, transfer functions and elliptic integrals \citep[see, e.g.,][]{cun-bar:1972,kar-etal:1992, vie:1993,sch:2006,rau-bla:1994,bec-don:2005,bro-loe:2005,bak-etal:2007,dex-ago:2009,vin-etal:2011,sch-stu:2013,cha-etal:2013,yan-wan:2014}. Our parallel code LSDplus uses reverse ray-tracing implemented by a direct numerical integration of the null geodesics.
  
Components of the four-momentum of a photon in the Kerr spacetime are given by  
\begin{gather}
   \label{CarterEQs}
   \begin{split}
     p^{r} = \dot{r} = s_r\Sigma^{-2} \sqrt{R_{\lambda,q}(r)}\,, \\
     p^{\theta} = \dot{\theta} = s_{\theta}\Sigma^{-2} \sqrt{\Theta_{\lambda,q}(\theta)}\,,\\
     p^{\phi} = \dot{\phi} = \Sigma^{-2} \Delta^{-1}
     \left[ 2ar + \lambda \left( \Sigma^{2} - 2r \right)  
\mathrm{cosec}^{2} \theta \right]\,, \\
     p^{t} = \dot{t} = \Sigma^{-2} \Delta^{-1} \left( \Sigma^{2} -
     2ar \lambda \right)\,,
   \end{split} 
	\end{gather} 
where dotted quantities denote differentiation with respect
to some affine parameter, and the sign pair $s_{r}$,$s_{\theta}$ describes orientation of radial and latitudinal evolution, respectively \citep[see, e.g.,][]{car:1968,mis-tho-whe:1973,chan:1983}.
Radial and latitudial effective potentials read  
\begin{gather}
   \label{eqn:2.1.3}
   \begin{split}
       R_{\lambda,q} \left( r \right) = \left[ \left( r^{2} + a^{2}
\right) - a \lambda \right] ^{2}
       - \Delta \left[ q - \left( \lambda - a \right) ^{2} \right]\,, \\
       \Theta_{\lambda,q} \left( \theta \right) = q + a^{2} \cos^{2}
\theta -\lambda^{2} \mathrm{cot}^{2} \theta\,.
     \end{split} 
		\end{gather} 
Here $\lambda, q$ are constants of motion related to the photons covariant angular and linear momenta. The LSDplus code performs a time-reverse integration of the package of null geodesics falling on a virtual detector of a distant observer with the screen resolution  1000 x 1000 pixels located at ($\theta_{obs}$, $r=1000$ M, $\varphi=0$) and traces the intersection of the geodesics with the disk surface corresponding to the emission events.

The code LSDCode enables modelling relativistic optical effects in the sky of the observer located anywhere above the event horizon of a Kerr black hole.
In the local reference frame related to the such general observer, the components of the four-momentum of photons with energy  normalized to one, falling on the pixel with coordinates $x$, $y$,  can be written as follows (see the right side of Figure \ref{fig_5}):
\begin{equation}
\label{for.mom}
k_{t} = -1, \,\,\, k_{r} = \sqrt{1 - x^{2} - y^{2}}, \,\,\,
k_{\theta} = - y, \,\,\, k_{\varphi} = - x. 
\end{equation}
Then one can obtain the coordinate covariant components of the four-momentum  by transforming the local components (\ref{for.mom}), using appropriate frame tetrads of one-form by relation
\begin{equation}\label{st.for.mom}
p_{\mu} = e_{\mu}^{\langle\alpha\rangle}k_{\langle\alpha\rangle}.
\end{equation}
The frame tetrad of one-form related to static observer in the Kerr spacetime is given as
\begin{eqnarray}
	e^{(t)}&=&\left\{ \sqrt{1-\frac{2r}{\Sigma}},0,0,\frac{2ar\sin^2\theta}{\sqrt{\Sigma^2-2r\Sigma)}}  \right\},\\
	e^{(r)}&=&\left\{0,\sqrt{\Sigma/\Delta},0,0 \right\},\label{LC9} \nonumber \\
	e^{(\theta)}&=&\left\{0,0,\sqrt{\Sigma},0 \right\}.\label{LC10}  \nonumber \\
	e^{(\varphi)}&=&\left\{ 0,0,0,\sqrt{\frac{\Delta\Sigma}{\Sigma-2r}}\sin\theta  \right\}. \nonumber
\end{eqnarray}
The constants of motion $\lambda(x,y)$ and $q(x,y)$ can be easily obtained by straightforward  calculation from the components of the four-momentum (\ref{st.for.mom}), using the relations \citep[see, e.g.,][]{chan:1983}
\begin{eqnarray}
\lambda &=& -\frac{p_{\phi}}{p_{t}}, \,\\
q^{2} &=& \left(\frac{p_{\theta}}{p_{t}}\right)^{2} + \left(\lambda\tan\left(\frac{\pi}{2} - \theta\right)\right)^{2} - a^{2}\cos^{2}\theta\,. \nonumber
\end{eqnarray}
However, in the investigated case of a distant static observer ($r_{obs}\to\infty$), when the rays  reaching the observer position are almost parallel (see Figure \ref{fig_5}), the relations between coordinates on the detector screen and constants of motion can be simply written as \citep[e.g.][]{cun-bar:1973}
\begin{equation}
x=-\frac{\lambda}{\sin\,\theta_{obs}}\,, \qquad y=\Theta_{\lambda,q} \left( \theta_{obs} \right)\,.
\end{equation}

In the event of detection, the constants $\lambda(x,y)$  and $q(x,y)$, together with the initial conditions (coordinates of observer and sign pair $s_{r}$,$s_{\theta}$) fully determinate the reverse temporal evolution of the zero geodesics of photons falling on a pixel of the detector screen with the coordinates  $x$,$y$.  The used Runge-Kutta method of the eighth order (Dorman-Prince method) \citep{pre-teu-vet-fla:2002} integrating the null geodesics reaches the relative accuracy of $10^{-15}$, which, in the case of a central black hole with stellar mass $M=5\,\mathrm{M}_{\odot}$, corresponds to the order of accuracy $10^{-11}$ meters on a radial coordinate. In order to determine the proper orientation of the radial and latitudinal component of the four-momentum (\ref{CarterEQs}), the code additionally analyses the positions of the radial and vertical turning point and sets the corresponding signs $s_{r}$ and $s_{\theta}$.  The integration of equations (\ref{CarterEQs}) by the the Runge-Kutta method of the eighth order proceeds naturally with adaptive step. However, the resulting null geodesic is then finally interpolated by the polynomials of the 3rd order for the chosen equidistant time step $\Delta T$, in the case of central black hole mass $M=5\,\mathrm{M}_{\odot}$, corresponding to $10^{-5}$ sec .

\subsection{Radiating surfaces and lightcurves}

The surfaces of the tori are modelled by the grid with resolution $15$ nodes in the radial direction and $75$ nodes in the azimuthal one. The grid contains the time-dependent information about coordinates, local intensity, and the four-velocity in the nodes. The time resolution of the surface of the tori necessarily corresponds to the time resolution of the interpolation steps of the geodesics package $\Delta T$.  The code LSDplus traces the intersections of the geodesics and linearly interpolated surface of a torus between triads of the nodes of the grid.  Assuming the normalized energy in the local observer's frame, the frequency ratio of the emitted and observed radiation from the torus can be expressed using projection of the four-momentum of a photon $p^{\mu}$ to the four-velocity of the surface of the torus $u_{\mu}^{sur}$ in the event of the emission as follows:
\begin{equation}
g = \frac{1}{p^{\mu}u_{\mu}^{sur}}.
\end{equation}
Radial and vertical components of the surface four-velocity are given by equations (\ref{rad.os.u},\ref{ver.os.u}). The remaining components $u_{\varphi}^{sur}$, $u_{t}^{sur}$ 
can be easily obtained using normalization condition $u^{\mu}u_{\mu}=-1$ together with the assumption of constant specific angular momentum \citep{vin-etal:2014}.
Then the instantaneous bolometric intensity detected by each pixel of the screen of a small virtual detector is calculated as
\begin{equation}
I_{obs}(t_{obs}) = I_{em}(t_{obs} - t_{delay})g^{4}, \label{bol_int}
\end{equation}
where $t_{delay}$ corresponds to the time delay (the change of a time coordinate) along the appropriate photon trajectory connecting the event of detection of a photon on a pixel with the event of emission on the surface of the torus. $I_{em}$ is local intensity on the surface in the comoving frame given by the equation (\ref{local_intensity}). The total instantaneous detected bolometric flux $F(t_{obs})$ is a sum of intensities (\ref{bol_int}) detected by individual pixels multiplied by the solid angle $\Delta \Pi$ subtended by the pixel in the observer sky   
\begin{equation}
\label{totalflux}
F(t_{obs}) = \sum^{1000}_{i = 1}\sum^{1000}_{j = 1}I_{ij}(t_{obs})\,\Delta \Pi, 
\end{equation}
where $i$ is the index of a pixel column and where $j$ is the index of a pixel row\footnote{We assume tiny angular size of the torus image in the observer sky and therefore constant $\Delta \Pi$.}. Since we are using relative units, we can simply set $\Delta \Pi=1$. The resulting lightcurves are generated in the time resolution $\Delta T$ corresponding to $20$ time samples per characteristic vertical oscillation period of the analyzed dual oscillation mode. 

\subsection{Power spectra}

In order to calculate power spectral densities (PSD), the resulting lightcurves are processed by fast Fourier transform (FFT).
The power spectral density at frequency $f_k=k/ (N \Delta t)$
is given as the square of the modulus
of the FFT of the signal as
\begin{equation}
PSD(f_k)= \left| \frac{1}{N}\sum_{j=0}^{N-1}F(t_j) \mathrm{exp}(2\pi ijk/N)\right|^{2}, \label{PSD}
\end{equation}
where $F(t_j)$ is the observed flux (\ref{totalflux}) at observation time $t_j = j\Delta t$ and $N$ is the total number of time samples in the lightcurve.

\subsection{Iron $K\alpha$ line profiles}
The fluorescent iron $K\alpha$ line consists of two components with FWHM $\approx 3.5\,eV$ and separation $\approx 13\,eV$ \cite[see][for details]{bas:1978}.
Considering the energy resolution $\Delta E=10\,eV$  of the simulation code, we approximated the rest
the rest iron $K\alpha$ line profile by Lorentzian peak with the scale factor $\gamma=20\,eV$ located on $E_0=6.4\,keV$. 
The instantaneous observed flux per pixel in the energy bin with the central energy $E_c$ is given as
\begin{equation}
\label{ebin_flux}
\Phi(t_{obs},E_c) = I_{em}(t_{obs} - t_{delay})g^{3}f(E_c/g,\gamma,E_0)\,\Delta \Pi,
\end{equation}
where Lorentzian function $f$ reads
\begin{equation}
f(x,\gamma,x_0)=\frac{1}{\pi\gamma[1+(\frac{x-x_0}{\gamma})^2]}\,.
\end{equation}
Then the observed instantaneous iron $K\alpha$ line profile is constructed by summing energy bin fluxes per pixels (\ref{ebin_flux})
over all pixels of the virtual detector in the given time sample. The final integrated line profile is obtained by summing instantaneous line profiles over all time samples. 

\section{Methods}
\label{sec:methods} 

\begin{figure*}
{\centering
\includegraphics[width=1.00\hsize]{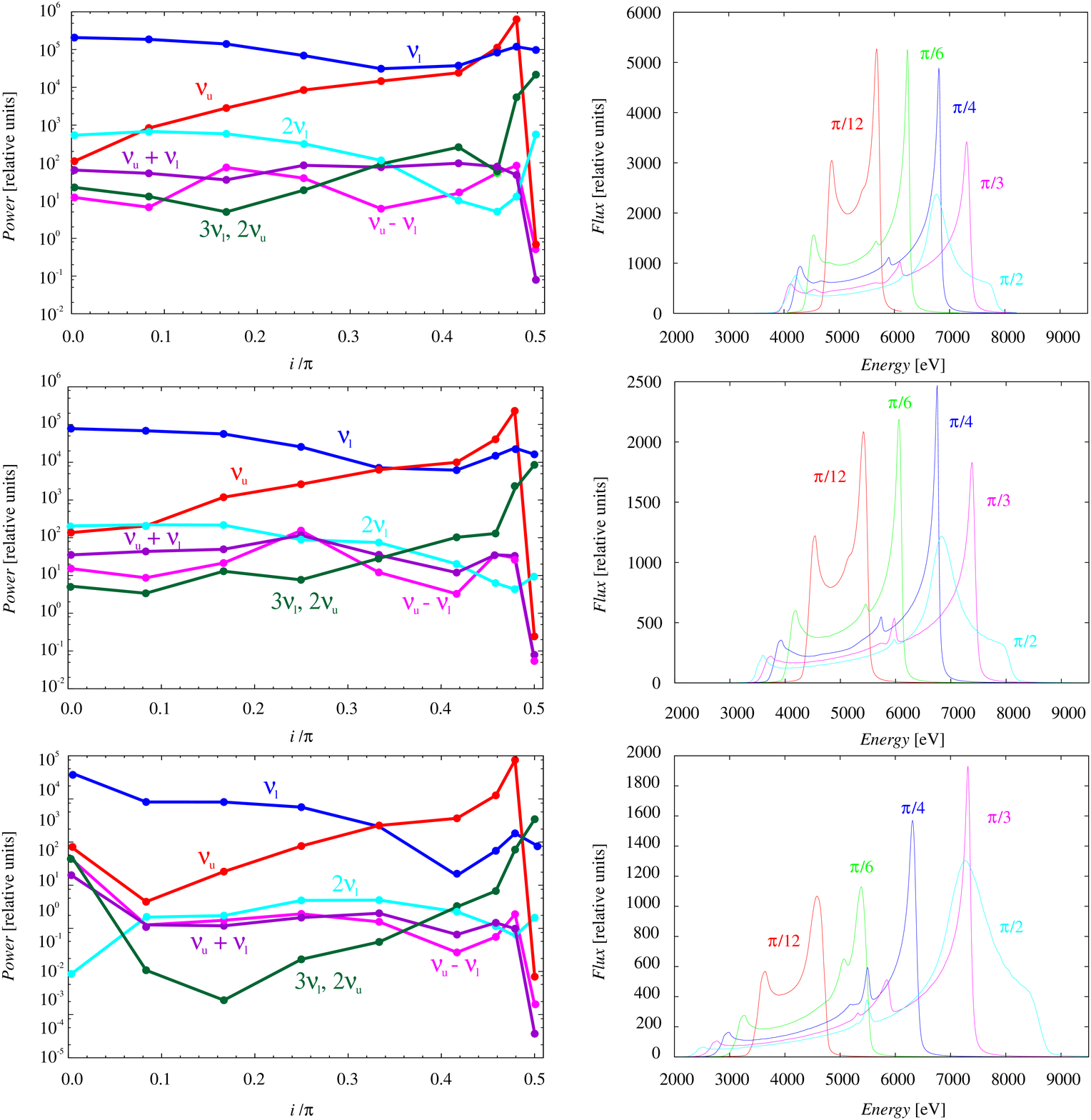}
}
\caption{\label{PSD1} The simulations outputs in the case of the ER axisymmetric dual mode. 
Left panels: The amplitudes of prominent PSD peaks (see Table \ref{tab:peaks}) as a function of distant observer inclination $i$ . Right panels: Iron $K\alpha$ line profiles constructed for $i \in \left(\frac{\pi}{12},\, \frac{\pi}{6},\,\frac{\pi}{4},\,\frac{\pi}{3},\,\frac{\pi}{2}\right)$.
The rows of the picture correspond to the cases of spin values $a=0$,\,$a=0.5$ and $a=0.96$, respectively (see Table \ref{tab:param1}).}
\end{figure*}

\begin{figure*}
{\centering
\includegraphics[width=1.00\hsize]{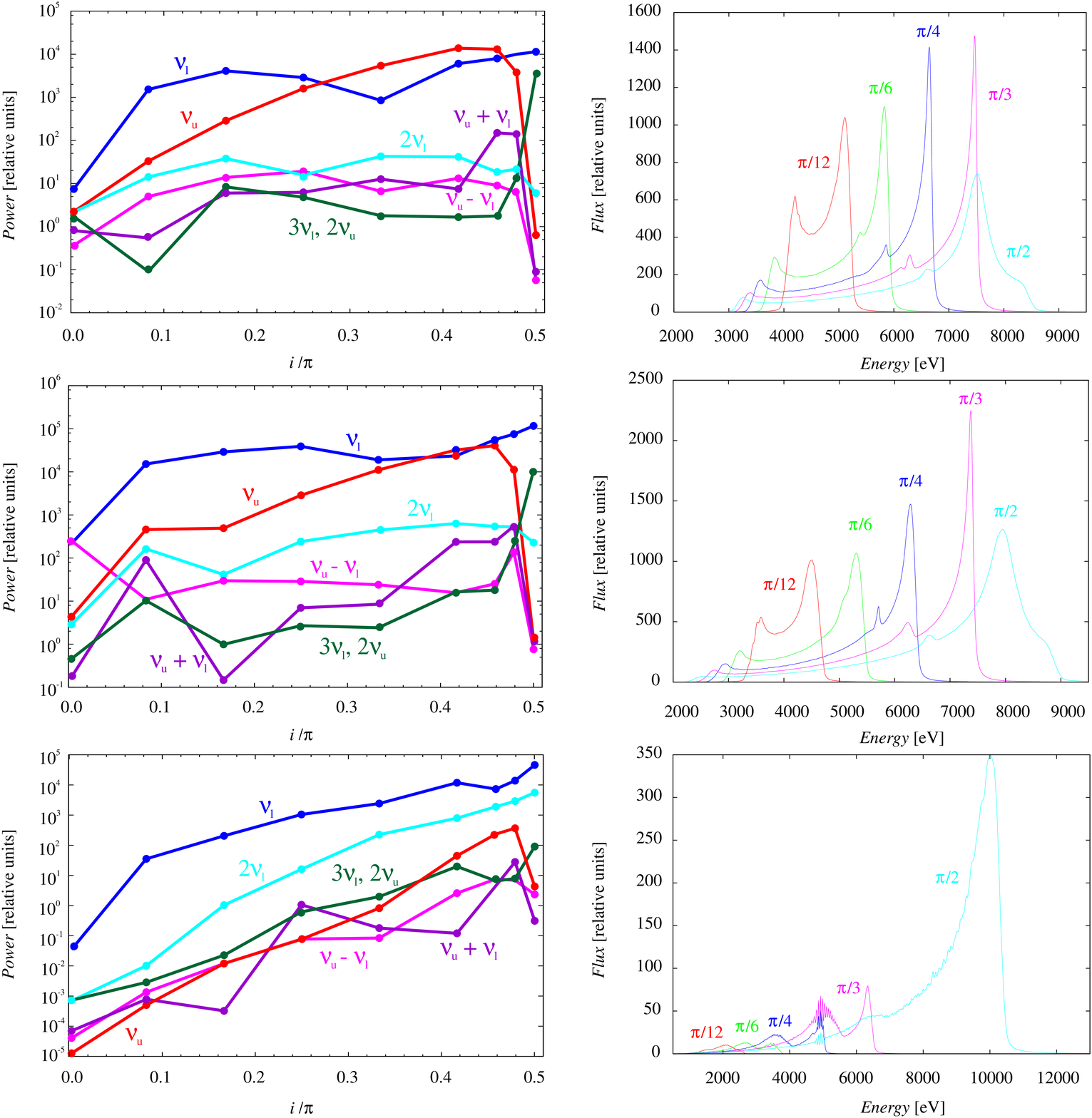}
}
\caption{\label{PSD2} The simulations outputs in the case of the RP-like non-axisymmetric dual mode. 
Left panels: The amplitudes of prominent PSD peaks (see Table \ref{tab:peaks}) as a function of distant observer inclination $i$ . Right panels: Iron $K\alpha$ line profiles constructed for $i \in \left(\frac{\pi}{12},\, \frac{\pi}{6},\,\frac{\pi}{4},\,\frac{\pi}{3},\,\frac{\pi}{2}\right)$.
The rows of the picture correspond to the cases of spin values $a=0$,\,$a=0.5$ and $a=0.96$, respectively (see Table \ref{tab:param2}).} 
\end{figure*}

Applying ray-tracing methods described in the previous section, we performed simulations of lightcurves of the above discussed model of a slender accretion torus oscillating in the dual mode regime. In order to obtain closed trajectories of the torus surface and corresponding closed lightcurves, we simulated behaviour of emitted radiation during three periods of vertical oscillation considering the fixed ratio of oscillation frequencies $\nu_{u}/\nu_{l}=3/2$. Then we calculated power spectra of obtained lightcurves by relation (\ref{PSD}).  We studied the impact of the central black hole spin using its three representative values $a \in \left(0,\,0.5,\, 0.96\right)$. The upper limit of the investigated spin value is almost the  highest one, for which the location of equilibrium torus centre $r_{0}$ can be found by the RP-like frequency relation (\ref{model2rel}). The impact of the observer inclination (polar angle) $i$ is analyzed using the following set of inclination values:
\begin{equation}
i \in \left(0.01,\,\frac{\pi}{12},\, \frac{\pi}{6},\,\frac{\pi}{4},\,\frac{\pi}{3},\,\frac{5}{12}\pi,\,\frac{5.5}{12}\pi,\,\frac{5.75}{12}\pi,\,\frac{\pi}{2}\right)\,. \nonumber
\end{equation}
We analyze the magnitude relations of five prominent PSD peaks corresponding  to two main dual mode oscillation frequencies, their higher harmonics and their sum or difference, particularly summarized in the Table \ref{tab:peaks}. 
\begin{table}[h!]
\begin{center}
\begin{tabular}{ll}
\hline
Frequency scaled in $\nu_l$ & Peak origin  \\
\hline
1/2 & $\nu_u-\nu_l$  \\
1 & $\nu_l=\left|\Omega_{r}/2\pi\right|$  \\
3/2 & $\nu_u=\left|\Omega_{\theta}/2\pi\right|$  \\
2 & $2\nu_l$  \\
5/2 & $\nu_u+\nu_l$  \\
3 & $3\nu_l$,\,$2\nu_u$  \\
\hline
\end{tabular}
\caption{\label{tab:peaks} {The most prominent peaks observed in power density spectra of simulated lightcurves.}}
\end{center}
\end{table}
Power spectra are calculated for all combination of investigated values of the spin and the inclination.  We plot magnitudes of these peaks as linearly interpolated functions of the inclination for all investigated values of the spin and for both investigated torus setups separately (see left panels of Figures \ref{PSD1} and \ref{PSD2}).

Moreover, the integrated iron $K\alpha$ line profiles emitted from torus surface are modelled for inclination values $i \in \left(\frac{\pi}{12},\, \frac{\pi}{6},\,\frac{\pi}{4},\,\frac{\pi}{3},\,\frac{\pi}{2}\right)$. The geometry of the torus optical projection is illustrated in Figure \ref{FSmaps} by the frequency shift maps on the virtual detector screen drawn in false colours related to the frequency shift values \footnote{The frequency shift maps displayed in Figure \ref{FSmaps} are not identically zoomed and positioned with respect to the whole observer sky.}. The maps clearly show that the simulation parameters are sufficient to distinguish the first three relativistic images. Also computed iron $K\alpha$ line profiles displays significant secondary blueshifted horns related to secondary (first indirect) relativistic images (see right panels of Figures \ref{PSD1},\ref{PSD2}).

\section{Results}
\label{sec:results} 

\subsection{Results for ER axisymmetric setup}

Table \ref{tab:param1} summarizes parameters of the slender torus oscillating in the pure epicyclic axisymmetric dual mode ($m_{r}=0$, $m_{\theta}=0$) for three investigated values of the spin. The radial coordinate of equilibrium torus centre $r_{0}$ given by relation (\ref{model1rel}) remains above the black hole photosphere\footnote{Kerr black hole photosphere -- a region of spherical unstable photon orbits reaches maximum extent in the equatorial plane between corotating and counter-rotating circular photon orbits, while it becomes infinitesimally thin on the polar axis \citep[see, e.g.,][for details]{teo:2003}} as well as the ergosphere for all such configurations. 
\begin{table}
\begin{center}
\begin{tabular}{llll}
\hline
Spin            & 0.0 & 0.5 & 0.96  \\
\hline
$\beta$ & 0.04 & 0.04 & 0.04  \\
Central radial coordinate $r_{0}$     & 10.80 &  7.92 &  4.35 \\
Torus radial extent     & 1.08 & 0.79 & 0.44 \\
Torus vertical extent  & 0.80 & 0.61 &  0.38  \\
Max. radial displacement &  0.27 & 0.20 &  0.11  \\
Max. vertical  displacement & 0.20 & 0.15 & 0.09  \\
\hline
\end{tabular}
\caption{\label{tab:param1} {Values of parameters describing the investigated slender torus model for ER axisymmetric setup.}}
\end{center}
\end{table}

The lightcurve waveforms are influenced by a complex interplay of the general relativistic frequency shift, the time variation of the emitting torus surface area and the time variation of the apparent torus area on the virtual detector screen \citep{bak-etal:2015}. All these effects strongly depend on the radial coordinate of the emission event, the inclination of the distant observer $i$ and the central black hole spin $a$. Nevertheless, the magnitude relations depicted on left panels of Figure \ref{PSD1} exhibit certain identical qualitative features for all investigated spin values. The pair of peaks corresponding to radial $\nu_l$ and vertical $\nu_u$ torus oscillation frequency is dominant on wide range of $i$. For this inclination range the examined ER axisymmetric setup predicts twin peaks HF QPOs frequency ratio equal to $3/2$ in accordance with expectations. The $\nu_l$ peak remains the most prominent for small and medium values of distant observer inclination $i$. The magnitude of the $\nu_u$ peak  grows with $i$ becoming the most prominent for relatively high inclinations, but it rapidly falls down for exact or almost exact equatorial observers. At the same time, the magnitude of the peak corresponding to $3\nu_l$,\,$2\nu_u$ rapidly grows. Therefore, in the case of such observers, the examined torus configuration predicts frequency ratio equal to $1/3$ for the pair of the most distinguishable HF QPO peaks. In the case of zero or moderate spin ($a=0$, $a=0.5$) and small inclination ($i \le \frac{\pi}{12}$), the magnitude of the $2\nu_l$ peak slightly exceeds the $\nu_l$ peak magnitude and the predicted twin peaks HF QPOs frequency ratio value is equal to $1/2$. As illustrated on the plots of magnitude relations (see left panels of Figure \ref{PSD1}), the spectral content of less distinct PSD peaks varies depending on values of $i$ and $a$. 

Like the power spectra behaviour, the Iron $K\alpha$ line profiles keep some identical qualitative features for all investigated spin values. Naturally, the energy span of line profiles is expanded and shifted down with decreasing $r_{0}$. The dependence of the height of the primary blueshifted horns  on $i$ is the only qualitative difference observable for different spin values (see right panels of Figure \ref{PSD1}).       

\subsection{Results for RP-like non-axisymmetric setup}

Table \ref{tab:param2} summarizes parameters of the slender torus oscillating in the RP-like non-axisymmetric dual mode ($m_{r}=-1$, $m_{\theta}=-2$) for three investigated values of the spin. In the case of zero or moderate spin ($a=0$, $a=0.5$), the radial coordinate of equilibrium torus centre $r_{0}$ given by relation (\ref{model2rel}) remains located above the black hole photosphere as well as the ergosphere, while in the case of high spin ($a=0.96$), the torus is located inside the ergopshere and therefore also deeply inside the photosphere. Moreover, in the case of high spin, the radial extent of the torus obtained as $r_{0}/10$ exceeds the location of the cusp of equipotential surfaces \citep{bla-etal:2006,str-sra:2009}. Such a torus configuration becomes unstable. Unfortunately, a surface area of the high spin stable configuration with maximum possible $\beta=7.0\,.\,10^{-5}$ is almost negligible and emitted flux is comparable to numerical error of the simulation. Therefore, we keep the torus radial extent equal to $r_{0}/10$ and choose the unstable high spin configuration with $\beta=0.002$.
\begin{table}
\begin{center}
\begin{tabular}{llll}
\hline
Spin            & 0.0 & 0.5 & 0.96  \\
\hline
$\beta$ & 0.02 & 0.02 & 0.002  \\
Central radial coordinate $r_{0}$     & 6.75 & 4.66 & 1.85  \\
Torus radial extent     & 0.68 & 0.47 & 0.18 \\
Torus vertical extent  &  0.27 & 0.18 & 0.02  \\
Max. radial displacement & 0.17 & 0.12 & 0.05  \\
Max. vertical displacement & 0.07 & 0.05 & 0.004  \\
\hline
\end{tabular}
\caption{\label{tab:param2} {Values of parameters describing the investigated slender torus model for RP-like non-axisymmetric setup.}}
\end{center}
\end{table}

In the case of zero or moderate spin ($a=0$, $a=0.5$), the qualitative picture of power spectra and iron $K\alpha$ line profiles behaviour is very similar to the case of the ER axisymmetric setup discussed in the previous section (see Figure \ref{PSD2}). The prediction for low inclination represents the main difference. In the Schwarzschild case, the pair of peaks corresponding to radial $\nu_l$ and vertical $\nu_u$ oscillation frequency remains dominant also for $i \le \frac{\pi}{12}$ predicting the twin peaks HF QPOs frequency ratio equal to $3/2$, as depicted on the top left panel of Figure \ref{PSD2}. In the case of $a=0.5$ and a very small inclination, the magnitude of the $\nu_u-\nu_l$ peak (instead of the $2\nu_l$ peak acting the same way in the case of the ER axisymmetric setup) slightly exceeds the $\nu_l$ peak magnitude, and the predicted twin peaks HF QPOs frequency ratio is equal to $1/2$ (see the middle left panel of Figure \ref{PSD2}). 

The unstable high spin configuration exhibits an entirely different picture, as shown in bottom panels of Figure \ref{PSD2}. In the whole range of inclination, the pair of dominant PSD peaks corresponds to the radial $\nu_l$ frequency and its second harmonic $2\nu_l$. Therefore, in the unstable high spin case, the examined RP-like non-axisymmetric setup surprisingly predicts the twin peaks HF QPOs frequency ratio value equal to $1/2$ only. The dramatic change of the optical projection properties is  documented by iron $K\alpha$ line profiles in the  bottom right panel of Figure \ref{PSD2}. The energy span is significantly shifted down, and both primary and secondary blueshifted horns have comparable heights. The computed profiles also display numerical instabilities caused by small emitting surface area of the unstable high spin torus configuration. The line profile related to $i=\frac{\pi}{2}$ exhibits exceptional behaviour, as its extremely wide energy span reaches $11\,keV$ and its primary blueshifted horn is absolutely dominant.  Corresponding frequency shift map in the bottom right panel of Figure \ref{FSmaps} also illustrates the equatorial optical projection of the torus located in the close vicinity of the Kerr black hole event horizon. It is clearly visible, that the angular size of both secondary and tertiary relativistic images exceeds the angular size of the primary image. Moreover, despite the significant gravitational redshift, the high Keplerian orbital velocity causes high Doppler blueshift of the left side of the torus projection corresponding to the extreme height of the primary blueshifted horn in the iron $K\alpha$ line profile.

\section{Conclusions and perspectives}
\label{sec:conc}

The aim of this article is to model twin peaks HF QPOs as a  spectral impact of isotropically radiating slender tori oscillating in a dual mode regime in the close vicinity of Kerr black holes. It was shown by \cite{vin-etal:2014} that significant fraction of the observed flux is regulated by the torus motion described by the lowest oscillation modes. Therefore, we examined two configurations of the dual oscillation regime based on the lowest radial and vertical oscillation modes with different azimuthal wave numbers. Appropriate frequency relations correspond to the two competing QPOs models, the epicyclic resonance (ER) HF QPOs model \citep{abr-klu:2001} and slightly modified relativistic precession (RP) QPOs model \citep{Ste-Vi:1999}. We model twin peaks HF QPOs by the pair of the most prominent peaks in the obtained power spectra. Our results show that independently of the spin, the ER axisymmetric setup yields power spectra with the pair of dominant PSD peaks corresponding to the frequencies of radial and vertical oscillation modes with proper ratio equal to $3/2$, except some special cases of very high or very low distant observer inclinations, where higher harmonics becomes prominent. The predictions of the RP-like non-axisymmetric setup are almost identical to the ER case for zero or moderate spin configurations.  An unstable high spin RP-like configuration with the torus located in the ergosphere exhibits dominant PSD peaks pair corresponding to the frequency of radial oscillation modes and its second harmonics in the whole range of inclinations and therefore it predicts constant frequency ratio equal to $1/2$. The entire change of the optical projection is also documented by the different iron $K\alpha$ line profiles with respect to the previous cases. 

The analysis presented in the article primarily focuses on relative ratios of the amplitudes of the most prominent frequency peaks in the modelled power spectra. However, the absolute fractional rms of the amplitudes of HF QPOs peaks in the detected signal depends not only on the amplitudes of perturbation $A_{r,\theta}$, but also on relations of individual components of the whole global source flux. In \cite{bak-etal:2014a} we defined an empirical model of global source flux, which mimics the so-called high steep power law (HSPL) state in GRS 1905+105, including steep spectrum and power-law dominated variability with an additional broad Lorentzian component at low frequencies {\citep{mcc-rem:2006}}. Then we used this background for the analysis of resolution of HF QPO peaks for a similar but simpler model of oscillating slender torus in the Schwarzschild geometry \citep{bur-etal:2004}, slowly passing the resonant orbit $r_0$. Considering the capabilities of the RXTE and LOFT instruments simulated by their response matrices, we have shown that the present available observational technology enables good detection of the pair of the most prominent peaks in such a modelled signal. Nevertheless, the behaviour of spectral content of less distinct PSD peaks remains probably the key task for the data analysis of future sensitive space observatory missions for X-ray timing, e.g. the proposed LOFT mission \citep[see][for details]{bak-etal:2014a,loft:2014,kar-etal:2014}.

Our simulation yields iron $K\alpha$ line profiles which are very different from the line profile integrated over the entire accretion disk. Therefore, the next related topic for future space X-ray missions can be sensitive frequency-resolved spectroscopy, which will be able to isolate the spectral component oscillating on the QPO frequency \citep[see, e.g.,][]{axe-don-hja:2014,rev-etal:1999}. Phase-resolved spectroscopy tracing the iron line profile changes throughout an oscillation cycle could be another potentially interesting diagnostic tool. Such idea has already been suggested for the case of low frequency QPOs \citep[e.g.,][]{ing-done:2012,tsa-but:2013}, but its application for the studied model of HF QPO  driven by slender torus oscillations will require a more detailed future study.

We assume an optically thick slender torus, and thus the influence of the torus opacity can be also subject for our future research. It is possible to generalize the examined slender torus model considering pressure effects on the mode frequencies and the torus shape \citep{str-sra:2009}. In the presented study, the location of the torus centre is empirically chosen using the observed twin peaks HF QPOs ratio. The resonant coupling of oscillation modes can prefer particular values of radial coordinate for amplification or excitation of QPOs \citep{hor:2008}. Therefore, further important improvement of the simulations can incorporate mode resonant coupling. Future implementation of improved models of non-slender tori into the used LSDplus code can be the next step towards more realistic results. A future detailed comparative study of spectral harmonic content can also be devoted to the possibility to distinguish between QPOs models based on either accretion tori oscillations or orbital motion of radiating blobs.

\begin{acknowledgements}
We acknowledge the Czech research grant GA\v{C}R~209/12/P740, Polish NCN UMO-2011/01/B/ST9/05439 grant and the internal grants of Silesian University in Opava IGS/11/2015 and SGS/11/2013.
\end{acknowledgements}

\bibliographystyle{aa}
\bibliography{OscillationTorusV3}

\end{document}